\documentclass[useAMS,usenatbib,usegraphicx]{mn2e}

\newcommand{\rp}{\mbox{$R_p$}}
\newcommand{\rs}{\mbox{$R_*$}}

\newcommand{\mpsini}{\mbox{$M_p \sin i$}}

\newcommand{\per}{\mbox{$P$}}
\newcommand{\w}{\mbox{$\omega$}}
\newcommand{\Tp}{\mbox{$T_p$}}
\newcommand{\apj}{ApJ}
\newcommand{\apjl}{ApJ}
\newcommand{\aap}{A\&A}
\newcommand{\mnras}{MNRAS}
\newcommand{\nat}{Nature}
\newcommand{\procspie}{Proc. SPIE}

\title[An ingress and a complete transit of HD~80606~b]{An ingress and a complete transit of HD~80606~b}
\author[Hidas et al.]{
  M. G. Hidas,$^{1,2,3}$\thanks{E-mail: mhidas@lcogt.net (MGH); ytsapras@lcogt.net (YT)} 
  Y. Tsapras,$^{1,4}$\footnotemark[1]
  D.~Mislis,$^{6}$
  A.~N.~Ramaprakash,$^{5}$
  S.~C.~C.~Barros,$^{8}$ 
\newauthor
  R. A. Street,$^{1,2}$
  J.~H.~M.~M~Schmitt,$^{6}$
  I. Steele,$^{7}$
  D. Pollacco,$^{8}$
  A. Ayiomamitis,$^{9}$
\newauthor
  J.~Antoniadis,$^{10}$
  A.~Nitsos,$^{10}$
  J.~H.~Seiradakis,$^{10}$
  S. Urakawa$^{11}$
\\
$^{1}$Las Cumbres Observatory Global Telescope, Goleta, CA 93117, USA \\
$^{2}$Department of Physics, University of California, Santa Barbara, CA 93106, USA \\
$^{3}$Sydney Institute for Astronomy, School of Physics, The University of Sydney, NSW 2006, Australia \\
$^{4}$School of Mathematical Sciences, Queen Mary, University of London, Mile End Road, London E1 4NS, England \\
$^{5}$Inter-University Centre for Astronomy and Astrophysics, Pune, India \\
$^{6}$Hamburger Sternwarte, Gojendergsweg 112, D-21029 Hamburg, Germany \\
$^{7}$Astrophysics Research Institute, Liverpool John Moores University, Liverpool CH41 1LD, UK \\
$^{8}$Astrophysics Research Centre, School of Mathematics and Physics, Queen's University Belfast, Belfast BT7 1NN, UK \\
$^{9}$Hellenic Astronomical Union, Athens, Greece \\
$^{10}$University of Thessaloniki, Department of Physics, Section of
Astrophysics, Astronomy and Mechanics, \\ 
GR-54124, Thessaloniki, Greece \\
$^{11}$Bisei Spaceguard Center, Japan Spaceguard Association, 1716-3 Ohrura, Bisei-cho, Ibara, Okayama 714-1411 Japan
}

\begin{document}

\date{Submitted 2010 Feb ??}

\pagerange{\pageref{firstpage}--\pageref{lastpage}} \pubyear{2010}

\maketitle

\label{firstpage}

\begin{abstract}
We have used four telescopes at different longitudes to obtain near-continuous
light curve coverage of the star HD~80606 as it was transited by its
$\sim$4-$M_{Jup}$ planet. The observations were performed during the predicted
transit windows around the $25^{th}$ of October 2008 and the $14^{th}$ of
February 2009. Our data set is unique in that it simultaneously constrains the
duration of the transit and the planet's period. Our Markov-Chain Monte Carlo
analysis of the light curves, combined with constraints from radial-velocity
data, yields system parameters consistent with previously reported values. We
find a planet-to-star radius ratio marginally smaller than previously reported,
corresponding to a planet radius of $\rp = 0.921 \pm 0.036 R_{Jup}$.
\end{abstract}


\section{Introduction}

An important class of the $\sim 400$ extra-solar planets known to date are the
so-called ``hot Jupiters'', orbiting a fraction of an AU from their host star.
Many of these have significantly larger radii than planets of similar mass in
our Solar System. This has been a challenge for theoretical
models of their structure. The intense radiation and tidal forces from the host
star are expected to play a significant role
\citep[e.g.][]{Bodenheimer2001,Burrows2007}.  For planets in highly eccentric
orbits, both of these factors vary greatly over the course of the orbit. Thus,
such planets provide interesting tests for models of the structure and dynamics
of planetary atmospheres \citep[e.g.][and references therein]{Irwin2008}. 

A planet in a 111-day orbit around the star HD~80606 was first detected in
radial velocity observations \citep{Naef2001}. The minimum mass (\mpsini) of the
planet is 3.9 times the mass of Jupiter.  HD~80606~b has the most eccentric
orbit of all the extra-solar planets known to date ($e = 0.93$). Infrared ($8
\mu$m) observations clearly show the rapid heating of the planet's atmosphere
during periastron passage \cite[herein referred to as
L09]{Laughlin2009}. Laughlin et al. also reported detection of a secondary
eclipse, implying the orbital inclination is close to 90 degrees, and
motivating efforts to observe a transit of the planet in front of the star.

In early 2009, several groups observed a transit egress in photometry
\citep{Moutou2009,Garcia-Melendo2009,Fossey2009} and spectroscopy
\citep{Moutou2009}. Analyses of these data, together with old and new
radial-velocity measurements provided the first constraints on the planet's
radius and actual mass, as well as its orbital parameters (\citealt{Pont2009},
herein referred to as P09; \citealt{Gillon2009}, G09). These analyses are
limited by the fact that the duration of the transit is not constrained,
which in turn increases uncertainties in the system
parameters. \citet[W09]{Winn2009} report multiple observations of the transit
ingress in June 2009, and combine their data with the earlier ingress
observations and Keck radial-velocity data to constrain the transit duration. 


In this paper we present previously unpublished multi-site photometric
observations of a complete transit of HD~80606~b on 14 February 2009, and an
ingress on 25 October 2008. We describe the observations and the resulting data
sets in section~\ref{sec:obsphot}. In section~\ref{sec:analysis} we describe
the methods we used to analyse our data, and in section~\ref{sec:results} we
report and discuss the results.


\section{Observations}
\label{sec:obsphot}

We observed HD~80606 with four telescopes, during predicted transit windows
around 25 October 2008 (herein referred to as Oct08), and 14 February 2009
(Feb09).\footnote{We obtained the predicted transit windows from {\it
http://transitsearch.org}.} As no transit or secondary eclipse had been
reported at the time, these windows were $\sim24$~hours long.

Where available, we observed through Bessell $B$ and $R$ filters, alternating
between them every few minutes. The one exception to this is noted below. In
all cases we deliberately defocused the telescope (often to the point where
the point-spread function became an annulus, but taking care that the target
and its binary companion do not blend). This allowed longer exposure times
without saturating the detector. More importantly, it meant that each star was
measured by a large number of pixels, reducing the effects of flat-fielding
errors and pointing drift on the photometry.

Besides those mentioned below, we also attempted observations at Bisei
Astronomical Observatory in Japan, Kryonerion Astronomical Station in Greece,
and an amateur observatory in Athens, Greece, but did not yield any useful data
due to adverse weather.

\subsection{Faulkes Telescope North, Hawaii}

The 2~m Faulkes Telescope North (FTN) --- part of the Las Cumbres Observatory
Global Telescope Network (LCOGT.net) --- is located on Haleakala in Hawaii. We
used FTN and a Merope camera with a 2k$\times$2k e2v CCD (binned $2\times
2$). Observations were conducted by the telescope's Robotic Control System,
alternating between sets of 37-sec exposures in $B$ and 20-sec exposures in
$R$. We only obtained useful data for the Feb09 transit. Images were processed
by our automated pipeline to remove the effects of bias, dark current and CCD
sensitivity variations.

\subsection{IUCAA Girawali Observatory, India}

Observations were carried out from IUCAA Girawali Observatory
\citep[IGO;][]{Das1999} and we obtained useful data during the Feb09 transit window.


We used the IUCAA Faint Object Spectrograph Camera \citep[IFOSC]{Gupta2002}
mounted on the direct Cassegrain focus of the IGO 2~m telescope. Exposure times
were 15--30~sec in $B$ and 4--10~sec in $R$. Images were corrected using
overscan strips and master bias frames followed by flat-fielding, using the
IFOSC photometry pipeline package developed under IRAF.

\subsection{Oskar-L\"{u}hning Telescope, Germany}
We observed HD~80606 on three nights in Feb09 with the 1.2~m
Oskar-L\"{u}hning Telescope (OLT) at Hamburg Observatory. We changed filters
after every 5 exposures, with exposure times of 55~s in $B$ and 35~s in $R$.
Images were bias-/dark-subtracted and flat-fielded using the {\it Figaro}
subroutine in the {\it Starlink} software package.

\subsection{Liverpool Telescope, Canary Islands}
Liverpool Telescope \citep[LT;][]{Steele2004} data were obtained on 2008
October 25 and 26 using the RISE frame transfer wide field camera
\citep{Steele2008} in $2\times 2$ binned mode, giving a pixel scale of
1.07~arcsec/pixel.  The filter employed consisted of 2~mm Schott KG5 bonded to
3~mm Schott OG515, giving half-maximum transmission from 515~nm to 695~nm
(i.e. roughly equivalent to a combined $V+R$ filter).  An exposure time of
1~sec was used, with effectively no dead time between exposures. All data were
dark-subtracted and flat-fielded using the ULTRACAM pipeline, which is
optimised for time series photometry. A master twilight flat was constructed
from a median of several hundred frames.

\begin{figure*}
\includegraphics[]{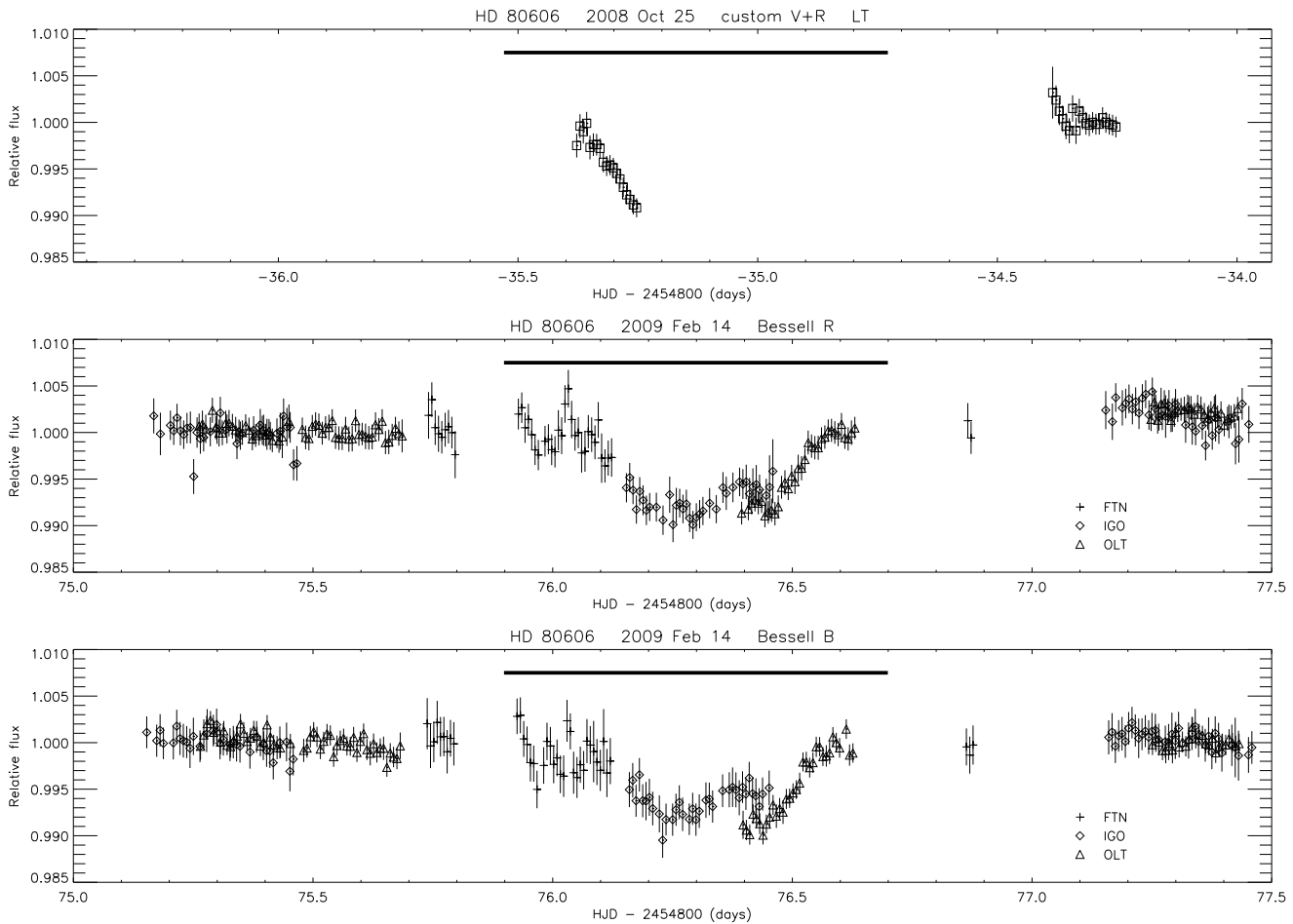}
\caption{The complete data set analysed in this paper: A transit ingress of
  HD~80706~b observed with the Liverpool telescope (LT, top panel), and the
  following transit observed by the Faulkes Telescope North (FTN), the IUCAA
  Girawali Observatory's 2m telescope (IGO), and the Oskar-L\"{u}hning
  Telescope (OLT). The middle and bottom panels show the same transit in the
  Bessell $R$ and $B$ bands, respectively. The horizontal bar across the top
  shows the section of data used in the fit. The error bars indicate the
  weights used in the fit, i.e. they include the statistical uncertainty as
  well as the red noise estimate ($\sigma_i$ and $\sigma_{r,s}$ in
  equation~\ref{equ:merit}).  \label{fig:lc}}
\end{figure*}

\subsection{Photometry} \label{subsec:photometry}
Starting with the calibrated images, we performed aperture photometry on all
data sets in the same manner, with the exception of the LT data.

For each instrument, a single master image was selected, and used as a
reference for all images obtained with that instrument. The coordinates of the
target and reference stars in the master image were obtained by convolving it
with a copy of the PSF from the same image, then running DAOFIND in IRAF on the
result. 

All other images were registered to the master using direct correlation with
the XREGISTER task. The master list of coordinates was then transformed into
the reference frame of each image. DAOPHOT was used to obtain photometry in a
range of apertures (slightly different for each instrument). The photometric
scatter was smallest in the largest aperture used for all data sets (radius
11'' for FTN, 8'' for IGO, and 10'' for OLT).

For the LT data, we also performed aperture photometry, trying out different
aperture radii, choosing the one that minimised the noise (8.6~arcsec).
Apertures were centred in each frame by cross-correlation of the image with a
Gaussian.

We measured the flux of HD~80606 relative to its binary companion HD~80607, as
it was the only reference star available in all data sets. Additional stars in
the fields of view of some of the instruments were significantly fainter and of
different colours, and were only used as check stars.

We prepared each light curve for fitting by taking the following steps.
\begin{enumerate}
\item We removed points more than $5\sigma$ from the median of the time series.
\item We normalised the flux values to a median of 1 for points well outside
  the transit.
\item We binned the light curve into 10-minute bins in order to reduce the
number of points to fit. We estimated uncertainties as the standard deviation
of the values within each bin, divided by the square root of the number of
points in the bin.\footnote{For bins containing only a single original point,
we estimated the uncertainty based on the nearest two bins with at least two
points.}
\end{enumerate}


\section{Analysis}
\label{sec:analysis}

The complete set of useful data we obtained is shown in Figure~\ref{fig:lc}. It
includes a transit ingress observed by the LT in October 2008, and --- between
the FTN, IGO and OLT data sets --- nearly continuous coverage of the transit in
February 2009 in two filters. 


\subsection{Light-curve model}
We modelled the HD~80606 system as a planet of radius \rp, orbiting a star of
radius \rs, with orbital period \per, semi-major axis $a$, eccentricity $e$,
and inclination $i$ (relative to our line of sight). The argument of periastron
is \w, and the planet passes there at time \Tp.

To obtain light curves for a given set of system parameters, we first solved
Kepler's equation using Newton-Raphson iteration, then calculated the planet's
position in its orbit, and its distance (in the plane of the sky) from the
centre of the star. Taking this as input, the analytical transit models of
\citet{MandelAgol2002} provided the relative fluxes expected for our physical
model. We assumed a quadratic limb-darkening law, with coefficients
interpolated from the tables of \citet{Claret2000}.\footnote{Based on the
stellar parameters reported by \citet{Naef2001}, we obtained the limb-darkening
coefficients $u_1=0.5043$, $u_2=0.2535$ in $R$ band, and $u_1=0.8827$,
$u_2=-0.0166$ in $B$ band. For the LT data set ($V+R$), we used the mean of the
coefficients for $V$ and $R$: $u_1=0.6117$, $u_2=0.1853$.}


\subsection{Fitting}\label{subsec:fitting}
We determined the parameters of the model using a Markov-Chain Monte Carlo
(MCMC) approach similar to that of \citet{TorresWinnHolman2008} and references
therein. Briefly, from a random starting point, small jumps in parameter space
are generated, and each new point is evaluated using a merit function
($\chi^2$). If the jump results in a lower $\chi^2$, it is executed (i.e. the
new point becomes the next ``link'' in the chain). Otherwise,
it is only executed with a probability $\exp(-\Delta\chi^2/2)$. If a jump is
not executed, the previous point is repeated in the chain. After a large
number of jumps, the result is a distribution of points in parameter space
that approximates the joint posterior probability density of all the
parameters given our data.

We used a merit function of the form:
\begin{equation}
\label{equ:merit}
\chi^2 = \sum_{i=1}^{N}{ \frac{(n_s f_i - m_i)^2}{\sigma_i^2+\sigma_{r,s}^2} }
       + \sum_{j=1}^{M}{ \frac{(v_j - c_j)^2}{\sigma_{c,j}^2} }
\end{equation}

The first sum is the usual chi-squared statistic and is related to the
probability of the data given the model. For
each of $N$ data points, $f_i$ is the measured relative flux, $m_i$ is the
model at that point, $\sigma_i$ is the statistical uncertainty in the
measurement, and $\sigma_{r,s}$ is an estimate of the correlated  noise in
each data set (see below). We also included a normalisation factor
$n_s$ (constant within each data set $s$) to allow the model to match the
out-of-transit relative flux.

The additional terms in the merit function represent constraints from
previously published measurements, as detailed in
Table~\ref{tab:constraints}. In each case a value calculated from our
model ($v_j$) is being compared to a measured value ($c_j$) with
corresponding uncertainty ($\sigma_{c,j}$). We included the time and
duration of the secondary eclipse from L09, as well as the period,
eccentricity and argument of periastron they derived from Keck
radial-velocity data. W09 note that the uncertainty in \w\ quoted by
L09 is likely to be underestimated. Indeed, both W09 and P09 analyse
larger data sets and find uncertainties in \w\ of
$\sim0.2$\degr. Therefore we adopt the value of \w\ from L09, but only
give it a weight corresponding to $\sigma_{c,\w} = 0.2$. We did not
include constraints from any other photometric observations in order
to obtain results as independent as possible from other analyses.

\begin{table}
  \caption{Measurements by \citet{Laughlin2009} included as constraints in the
  MCMC merit function. The three columns correspond to $v_j$, $c_j$ and
  $\sigma_{c,j}$, respectively, in equation~\ref{equ:merit}. The last two rows
  refer to the epoch and duration ($2^{nd}$ to $3^{rd}$ contact) of the
  secondary eclipse.} \label{tab:constraints}
  \begin{tabular}{@{}lrl}
    \hline
    Parameter & Value & Uncertainty \\
    \hline
    Orbital period, \per\ (days) &    111.4277 & 0.0032 \\
    Orbital eccentricity, $e$   &      0.9327 & 0.0023 \\
    Argument of periastron, \w\ (deg)  &    300.5    & 0.2\,$^a$ \\
    Epoch of mid-eclipse (HJD)  & 2454424.736 & 0.003  \\
    Total-eclipse duration (days) &       0.070 & 0.009  \\
    \hline
  \end{tabular}
  $^a$ This is our conservative estimate of the uncertainty (see section~\ref{subsec:fitting}). The value reported by \citet{Laughlin2009} is \w$= 300.4977 \pm 0.0045$\degr.
\end{table}

During the MCMC simulation we varied the following parameters: $a/\rs$, \per,
\Tp, $e'$ (see below), \w, $b = a/\rs \cos i$ (proportional to the transit
impact parameter), $\rp/\rs$, and a normalisation factor for each
instrument/filter combination ($n_s$). A total of 14 parameters were varied. In
order to avoid strong correlations which restrict the movement of the MCMC
chain through parameter space, we used combinations of some of the physical
parameters that are more directly constrained by the data (e.g. the radius
ratio instead of the absolute radii). When eccentricity was included as one of
the varied parameters, correlations caused different MCMC chains to converge to
different locations in (\per, $e$, \w) space. We found that most of the
variation in $e$ over all chains could be described as a linear function of
\per and \w. Thus, we replaced $e$ with $e' = e - k_0\per - k_1\w$, where $k_0
= 0.035$ and $k_1 = -0.0035$ were determined from the previous fit. This
resulted in the MCMC simulation moving more freely through parameter space and
a more continuous distribution of points when the chains were combined.

We generated jumps by adding a Gaussian-distributed
random value to one of the varied parameters (randomly chosen) at each
iteration. We tuned the jump sizes for each parameter so that approximately
25--50\% of jumps were executed.

\subsection{Correlated noise}

Each of our data sets contains some noise that is correlated in time due to
gradual variations in atmospheric or instrumental factors. The {\it internal}
uncertainties we estimated for each data point (as described in
section~\ref{subsec:photometry}) do not measure this {\it red noise}, yet it
can significantly affect the results of our analysis \citep{Pont2006}. To
account for the effect of red noise, we reduced the weight of each point in the
merit function (equation~\ref{equ:merit}) with an additional noise term,
$\sigma_{r,s}$, which we estimated as follows. We first performed an initial
fit setting $\sigma_{r,s}$ to an optimistic 0.0005 for all data sets. For each
data set $s$ (of $N_s$ points), we then found the value of $\sigma_{r,s}$ for
which the reduced chi-squared
\begin{equation}
\label{equ:rchisq}
\chi^2_r  = \frac{1}{N_s-8} 
            \sum_{i=1}^{N_s}{ \frac{(f_i-m_i)^2}{\sigma_i^2+\sigma_{r,s}^2} }
\end{equation}
was equal to unity.\footnote{The effective number of parameters for this
chi-squared is 8, consisting of the 7 physical parameters and one normalisation
factor $n_s$ for the given data set.}  We then repeated the fit, and checked
that the values of $\sigma_{r,s}$ were self-consistent. The final values used
were
LT: 0.001, 
FTN($R$): 0.0016, 
IGO($R$): 0.0015,
OLT($R$): 0.0012,
FTN($B$): 0.0019,
IGO($B$): 0.0017,
OLT($B$): 0.001.
The average internal uncertainty of data used in the fit was only 0.0006, so
--- as is generally the case in such analyses --- red noise is the main factor
limiting the accuracy of our results.

\begin{figure*}
\includegraphics[]{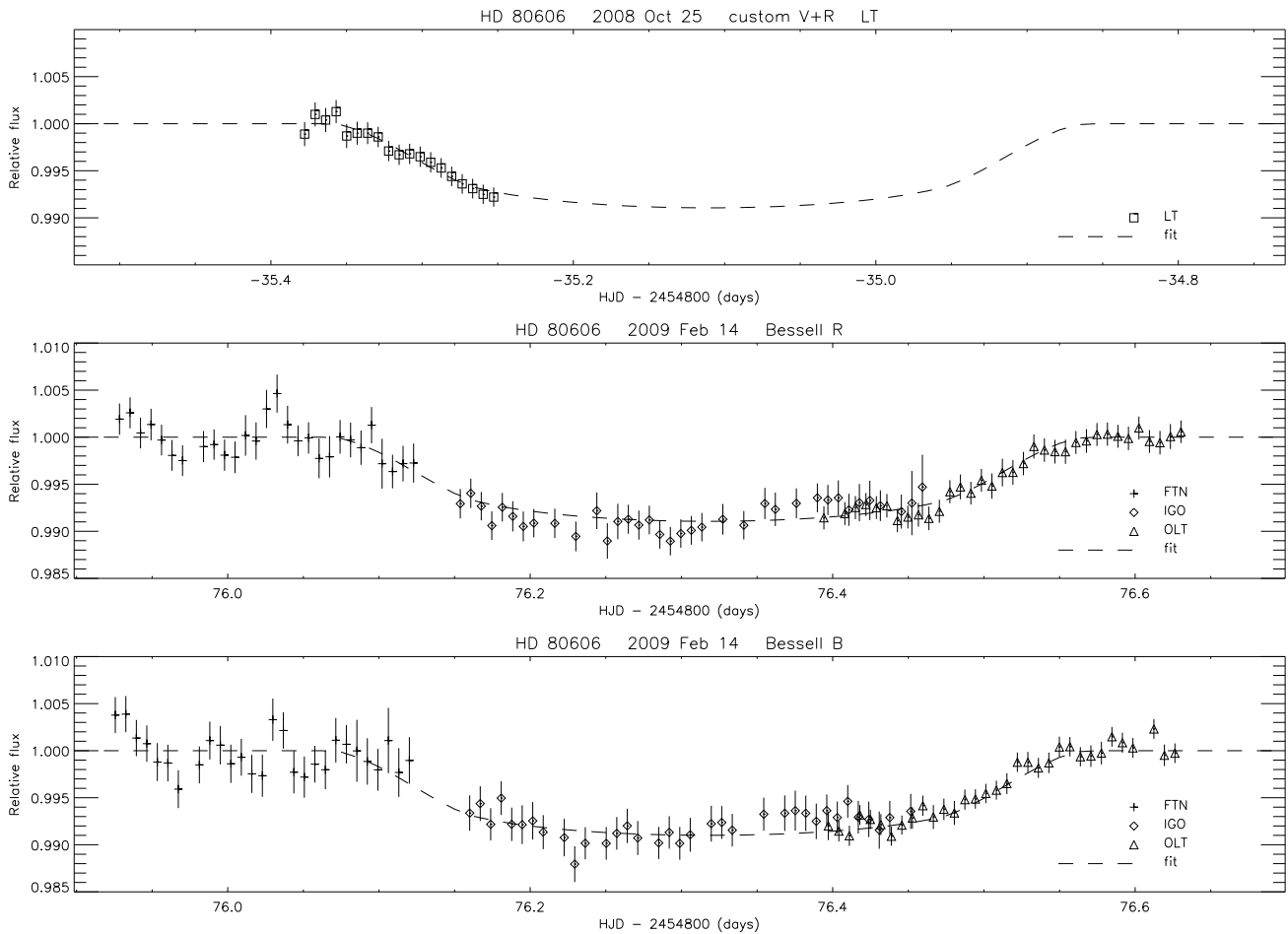}
\caption{Close-up of the data used in the fit, along with the best-fit model
  obtained from MCMC simulations. Symbols and error bars are the same as in
  Figure~\ref{fig:lc}, only the horizontal scale is different.
  \label{fig:fit}}
\end{figure*}

\begin{table*}
\begin{minipage}{126mm}
  \caption{Physical parameters of the HD~80606~b system obtained from the MCMC fit.} \label{tab:results}
  \begin{tabular}{lcc}
    \hline
    Parameter & Median & Uncertainty \\
    \hline
    Scaled semi-major axis, $a/\rs$  &  101.2     &  -5.5, +6.9 \\
    Orbital period, \per\ (days)     &  111.4273  &  $\pm$0.0031 \\
    Orbital eccentricity, $e$        &    0.93369 &  $\pm$0.00068 \\
    Epoch of periastron, \Tp\ (HJD)  &2454424.8575&  $\pm$0.0040 \\
    Argument of periastron, \w\ (deg)&  300.53    &  $\pm$0.19 \\
    Orbital inclination, $i$ (deg)   &   89.341   & -0.063, +0.073\\
    Planet/star radius ratio,$\rp/\rs$ &  0.0967  & -0.0035, +0.0032 \\
    Planet radius, $\rp$\,$^a$ ($R_{Jup}$) & 0.921   &  $\pm$0.036 \\
    \hline
    Epoch of mid-transit (HJD) &  2454876.3173 &   $\pm$0.0036 \\
    Complete transit duration ($t_4-t_1$, hours) &   12.14  &   $\pm$0.36 \\
    Ingress/egress duration ($t_2-t_1$, hours)  &   2.33 &   $\pm$0.37 \\
    \hline
  \end{tabular} 
  \\
  $^a$ Using $\rs = 0.978 \pm 0.015 R_{\sun}$ \citep{Pont2009}.
\end{minipage}
\end{table*}


\section{Results and Discussion}
\label{sec:results}

For our final fit, we created 25 
independent chains of 42,000 
steps each, and discarded the first 2,000 
steps of each chain to minimise the effect of the (randomly chosen) starting
points on our results. Combining the remaining steps of all the chains, the
resulting distribution of parameter values approximates their joint probability
density. For each parameter we report the median of this distribution as the
``best-fit'' value,\footnote{As \citet{Pont2009} point out, the set of
parameters thus obtained does not actually correspond to the MCMC step with the
lowest $\chi^2$ value. This is because each parameter's median value comes from
a different step in the chain.}  and the boundaries of the central 68\% as the
1-$\sigma$ uncertainties. 

The results thus obtained are shown in Table~\ref{tab:results} and the fit
itself is compared to the data in Figure~\ref{fig:fit}. The reduced chi-squared
for this fit is 0.63, which suggests we have slightly overestimated the
uncertainties in our data. Thus the uncertainties in our results are somewhat
conservative. We also note that there are unavoidable correlations between some
parameters. Besides the \per-$e$-\w\ correlation mentioned in
section~\ref{subsec:fitting}, the strongest of these are between $a/\rs$, $b$
and $\rp/\rs$. There are also significant correlations between \Tp\ and \w, and
also between $\rp/\rs$ and the normalisation factor for the IGO data. The
uncertainties in results include the effects of all these correlations.

Since we used the values of the period (\per), eccentricity ($e$) and argument
of periastron (\w) from L09 as direct constraints in the fit, we obtained
results that agree with those values. We did find a larger uncertainty for \w,
but a more precise value for the eccentricity.

Our results also agree well with the parameters reported by P09, G09 and
W09. The only significant exception to this is the period, where our result is
$3\sigma$ smaller than the other analyses. We note, however, that we have
essentially forced our simulations to agree with the period reported by L09. If
we repeat our fit without the direct constraints on \per, $e$ and \w\ (see
section~\ref{subsec:fitting}), the best-fit period is almost unchanged, but the
uncertainty in its value is four times greater. This means our photometric data
are in fact consistent with the period reported by P09, G09 and W09.

We also find a planet-to-star radius ratio that is marginally smaller (by
$1.8\sigma$) than the values reported by W09 and G09. A possible cause for this
is the nature of our data from the IGO, which contains the majority of the
in-transit points, but no out-of-transit points on the same night. This means
the normalisation of those data is not strongly constrained, and leads to the
correlation noted above. In addition, the IGO data appear to show a significant
systematic trend in the middle of the transit, which we were unable to remove
by de-correlation against any physical parametrers of the observations (such as
airmass or position on the CCD).


\section{Conclusion}
\label{sec:conclusion}

We have obtained multi-site observations of a transit ingress and a complete
transit of HD~80606~b across its host star. We analysed these data
independently of any other photometric data, and found system parameters
consistent with previously reported values.

These observations were made using four telescopes at different sites.
This allowed us to obtain near-continuous coverage of this 12-hour
event. However, the differences between the instruments, telescopes
and time-allocation procedures were, to an extent, limitations on our
ability to obtain a uniform data set. In the near future, the
completion of LCOGT's network of 1m robotic telescopes \citep{Brown2010} will greatly facilitate observations of this kind, providing near-identical
instrumentation at a number of sites, under the control of a flexible,
central scheduling system.



\section*{Acknowledgments}

We are grateful to Greg Laughlin for pointing out the opportunity to observe
this transit. We thank Tom Marsh for the use of the ULTRACAM pipeline.


\bsp

\label{lastpage}


\begin{thebibliography}{}

\bibitem[Das et al.(1999)]{Das1999} Das, H.~K., Menon, S.~M., 
Paranjpye, A., \& Tandon, S.~N.\ 1999, Bulletin of the Astronomical Society of India, 27, 609 

\bibitem[Bodenheimer et al.(2001)]{Bodenheimer2001} Bodenheimer, P., 
  Lin, D.~N.~C., \& Mardling, R.~A.\ 2001, \apj, 548, 466 


\bibitem[Burrows et al.(2007)]{Burrows2007} Burrows, A., Hubeny, 
  I., Budaj, J., \& Hubbard, W.~B.\ 2007, \apj, 661, 502 

\bibitem[Claret(2000)]{Claret2000} Claret, A.\ 2000, \aap, 363, 1081 



\bibitem[Fossey et al.(2009)]{Fossey2009} Fossey, S.~J., Waldmann, I.~P., \&
  Kipping, D.~M.\ 2009, \mnras, 396, L16

\bibitem[Garcia-Melendo \& McCullough(2009)]{Garcia-Melendo2009}
  Garcia-Melendo, E., \& McCullough, P.~R.\ 2009, \apj, 698, 558


\bibitem[Gillon(2009)]{Gillon2009} Gillon, M.\ 2009, submitted to MNRAS
  (arXiv:0906.4904)

\bibitem[Gupta(2002)]{Gupta2002} Gupta et al. 2002, BASI, 30, 785  


\bibitem[Irwin et al.(2008)]{Irwin2008} Irwin, J., et al.\ 2008, \apj , 681,
  636

\bibitem[Laughlin et al.(2009)]{Laughlin2009} Laughlin, G., Deming, D.,
  Langton, J., Kasen, D., Vogt, S., Butler, P., Rivera, E., \& Meschiari, S.\
  2009, \nat, 457, 562


\bibitem[Mandel \& Agol(2002)]{MandelAgol2002} Mandel, K., \& Agol, E.\ 2002,
  \apjl, 580, L171

\bibitem[Moutou et al.(2009)]{Moutou2009} Moutou, C., et al.\ 2009, \aap, 498,
  L5

\bibitem[Naef et al.(2001)]{Naef2001} Naef, D., et al.\ 2001, \aap, 375, L27

\bibitem[Pont et al.(2006)]{Pont2006} Pont, F., Zucker, S., \& Queloz, D.\
  2006, \mnras, 373, 231

\bibitem[Pont et al.(2009)]{Pont2009} Pont, F., et al.\ 2009, \aap, 502, 695 

\bibitem[Steele et al.(2004)]{Steele2004} Steele, I.~A., et al.\ 2004,
  \procspie, 5489, 679

\bibitem[Steele et al.(2008)]{Steele2008} Steele, I.~A., Bates, S.~D., Gibson,
  N., Keenan, F., Meaburn, J., Mottram, C.~J., Pollacco, D., \& Todd, I.\ 2008,
  \procspie, 7014, 217



\bibitem[Torres et al.(2008)]{TorresWinnHolman2008} Torres, G., Winn, J.~N., \&
  Holman, M.~J.\ 2008, \apj, 677, 1324

\bibitem[Winn et al.(2009)]{Winn2009} Winn, J.~N., et al.\ 2009, \apj, 703,
  2091

\bibitem[Brown et al. (2010)]{Brown2010} Brown, T.~M., et al.\ 2010, Bulletin of the American Astronomical Society, 41, 401-+


\end{thebibliography}
\end{document}